# Relativity of Time in Earthquake Physics

A.V. Guglielmi, O.D. Zotov

*Schmidt Institute of Physics of the Earth, Russian Academy of Sciences; Bol'shaya Gruzinskaya str., 10, bld. 1, Moscow, 123242 Russia; guglielmi@mail.ru (A.G.), ozotov@inbox.ru (O.Z.)*

**Abstract**

In this paper we develop an understanding of the proper time of the Tohoku earthquake source. The paper is dedicated to the 120th anniversary of Einstein's theory of relativity, but the dedication is symbolic, since we are investigating a purely non-relativistic geophysical object. We still found it possible to borrow the terms "time relativity" and "proper time" from the theory of relativity. The concept of the proper time of the source is developed by us within the framework of the phenomenological theory of earthquakes. The paper describes a procedure for measuring proper time using observation data of foreshocks and aftershocks. We synchronized the imaginary "underground clock" that counted the proper time and the clock that showed world time. The idea of a hypothetical underground clock turned out to be effective. We have shown that the proper time of the source flows unevenly relative to the flow of world time. Two phases of the evolution of the source before the main shock of the earthquake were discovered. This complements the picture of three-phase relaxation of the source after the main shock, which we established earlier. The uneven flow of proper time relative to world time indicates the non-stationarity of the source parameters. The overall conclusion is that the concept of the proper time of the source has enriched the possibilities of experimental study of earthquakes.

## 1. Introduction

This paper completes a short series of our publications [1, 2] dedicated to the 120th anniversary of Einstein's theory of relativity. The dedication is symbolic, since we are investigating a purely non-relativistic geophysical object – the source of an earthquake. However, we still found it possible to borrow the terms "time relativity" and "proper time" from the theory of relativity.

The logic of the development of the phenomenological theory of earthquakes led us to the idea of the proper time of the earthquake source. The phenomenological theory is based on the idea of the source as a dynamic system. The state of the system after the main shock of an earthquake is described by the so-called deactivation coefficient $\sigma(t)$ (see reviews [3–5] by a group of authors who proposed a phenomenological theory of earthquakes). Proper time $\tau$, different from universal time, is calculated using the formula

$$\tau = \int_0^t \sigma(t')dt'. \qquad (1)$$

The source as a dynamic system is not accessible for direct study using the "input-output" scheme. Figuratively, it can be imagined as a "black box without an entrance". We judge the dynamics of the source by the sequence of aftershocks, which we consider as signals from the system's output.

To synchronize the abstract clock that keeps its own time and the clock that shows world time, a relationship is postulated between the aftershock frequency $n(t)$, determined experimentally, and the deactivation coefficient $\sigma(t)$. The postulate is formulated on the basis of the following plausible reasoning [6, 7]. According to Omori's law [8], the frequency of aftershocks depends nonlinearly (hyperbolically) on time: $n \propto 1/t$. Let's perform linearization by changing the variable $n \to g = 1/n \propto t$. Now it is natural to base the elementary theory of aftershocks on the axiom $\sigma = dg/dt$. It is equivalent to the differential equation

$$\frac{dn}{dt} + \sigma n^2 = 0, \qquad (2)$$

the solution of which has the form

$$n(t) = \frac{n_0}{1 + n_0 \tau(t)}, \qquad (3)$$

where $n_0 = n(0)$ is the initial condition. Thus, in the phenomenological theory, the frequency of aftershocks decreases hyperbolically with respect to the proper time of the source, and not at all with respect to world time according to Omori's law. The question of the applicability of Omori's law became the subject of a simple experimental test. Namely, the law is satisfied if and only if $\sigma = \text{const}$.

Let us emphasize two differences between the phenomenological theory and Omori's theory [8] and its modifications (see, for example, [9–16]). Firstly, the object of study is the earthquake source, and not the aftershocks themselves. Second, unlike Omori, we make no prior assumptions about the dependence of aftershock frequency on universal time. The answer to the question about the law of attenuation of aftershocks over world time is sought experimentally by solving the inverse problem of the source (IPS), which is formulated as follows: determine the coefficient of deactivation of the source based on the data on the frequency of aftershocks. The correct solution of the IPS has the form

$$\sigma(t) = \frac{d}{dt} \langle g(t) \rangle, \qquad (4)$$

where the angle brackets denote the optimal smoothing of the auxiliary function $g(t)$.

The continuous function $n(t)$ is formed by averaging a discrete sequence of aftershocks over small intervals of universal time, each of which contains a sufficiently large number of aftershocks. The uncertainty of the averaging procedure prompted us to look for an alternative way of introducing the proper time of the source, which directly takes into account the discreteness of the tremors. Our new approach [1, 2] to the concept of the proper time may seem strange, but it has proven to be quite effective. We looked at aftershocks not only as a object of investigation, but also as a kind of ticking of a hypothetical "underground clock" that counts down the source's own time. Simply put, we agreed to consider the excitation of the next underground shock as evidence that a unit of the proper time of the source has passed since the moment of the previous underground shock.

The idea of a hypothetical underground clock that keeps time through underground shocks has proven fruitful in the analysis of the aftershocks of the Tohoku earthquake [2]. The main shock with a magnitude of $M = 9.1$ occurred on March 11, 2011. Figure 1 shows the epicenters of foreshocks and aftershocks. In the 200-day intervals before and after the main shock, 166 foreshocks and 4537 aftershocks occurred. In this paper, we focus on the foreshocks of the Tohoku earthquake.

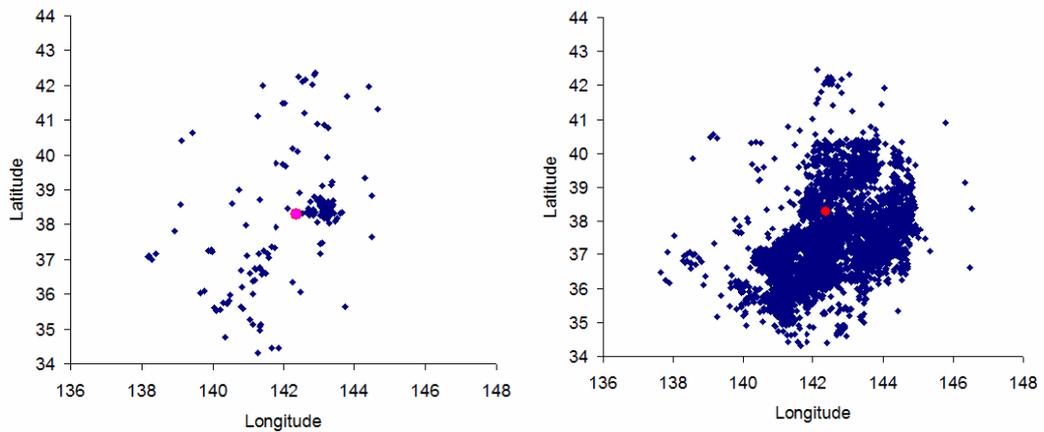

**Fig. 1**. Foreshock (left) and aftershock (right) epicenters 200 days before and 200 days after the mainshock of the Tohoku earthquake. The red circles indicate the location of the epicenter of the mainshock. The figures are based on data from the USGS/NEIC catalog (https://earthquake.usgs.gov).

## 2. Foreshocks of the Tohoku earthquake

Figure 2 gives a general idea of the foreshock and aftershock dynamics of the Tohoku earthquake. The figure is based on data from the USGS/NEIC catalog. Over a 400-day interval, 166 foreshocks and 4537 aftershocks were recorded.

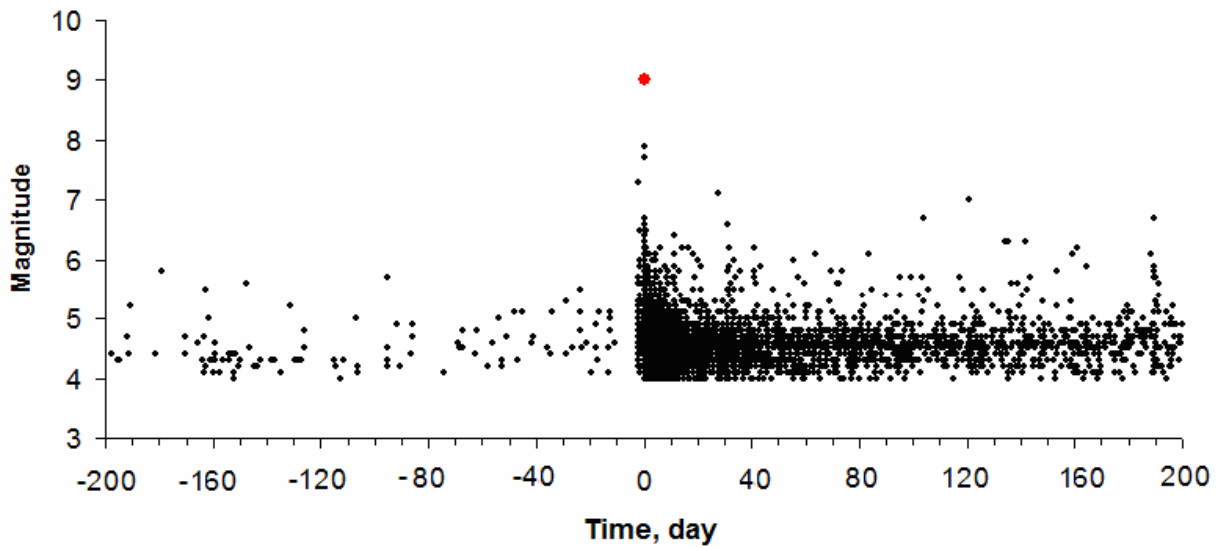

**Fig. 2**. Dependence of the magnitude of foreshocks and aftershocks on world time. The red dot indicates the magnitude of the main shock of the Tohoku earthquake.

Our idea is to use the tremors as the ticking of an imaginary underground clock, counting down the source proper time. We want to synchronize the underground clock and the clock showing world time. To do this, we number the sequence of foreshocks, main shock and aftershocks with numbers $k$ = 1, 2, 3, .., 4704. Using the USGS/NEIC catalog, we determine the world time $t_k$ for each $k$. We will plot points $t_k$ on the coordinate plane $(x, t)$. Here the letter $x$ denotes the proper time of the source (from the word "хронос", or χρόνος). The result is shown in Figure 3.

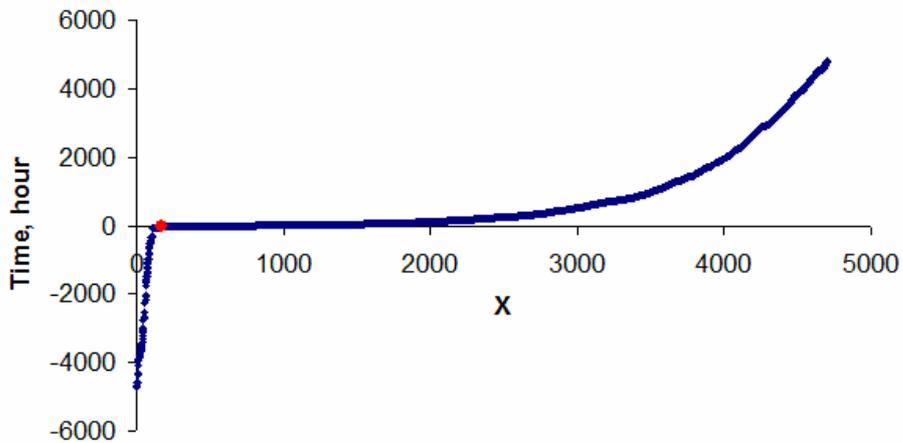

**Fig. 3**. Dependence of the occurrence times of foreshocks, mainshock and aftershocks according to universal time (vertical axis) on the proper time of the Tohoku earthquake source (horizontal axis). The moment of the main shock is marked by a red dot.

The result was surprising. The aftershock $t_k$ points merged into a perfect curve, almost indistinguishable from an exponential curve. But we have already carefully analyzed the aftershocks in our work [2]. Let's now focus on foreshocks.

In Figure 3, the foreshocks correspond to x values from 1 to 166. The points $t_k$ are approximately located along a straight line $t(x) = 32x - 4169$. However, a more detailed analysis of foreshock dynamics revealed a different picture. The point is this.

51 hours before the main shock, a powerful foreshock with a magnitude of $M$ = 7.3 occurred. It was followed by 63 foreshocks, the dynamics of which were radically different from the dynamics of the preceding 103 foreshocks. We analyzed the two foreshock sets separately and presented them in enlarged scale in Figures 4 and 5.

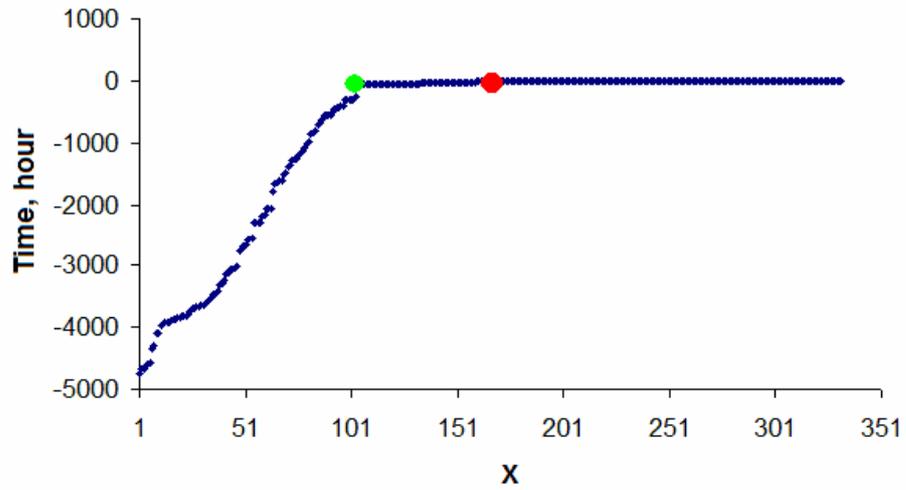

**Fig. 4**. The same as in Figure 3 on an enlarged scale. Shown are 166 foreshocks, the mainshock (red dot), and 166 aftershocks. The green dot marks the moment of a distinct change in the foreshock evolution regime.

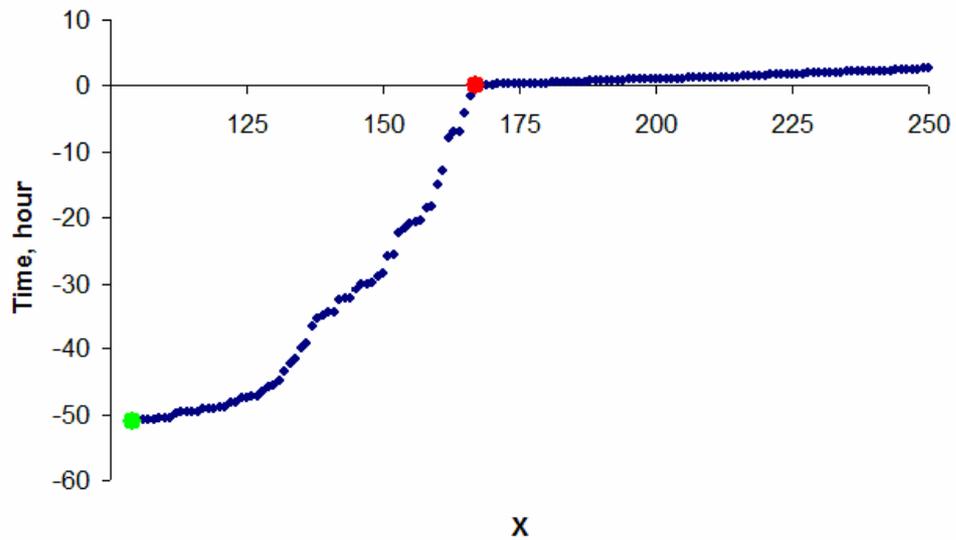

**Fig. 5**. The same as in Fig. 3 on the time interval from $x = 104$ to $x = 250$.

In Figure 4 we see that foreshocks one through one hundred and three are approximated by a straight line

$$t(x) = 46.5x - 485. \qquad (5)$$

The coefficient of determination is 0.98. In contrast, the next 63 experimental points are located along the exponential curve

$$t(x) \propto \exp(0.065x), \quad (6)$$

and the coefficient of determination is also quite high (0.97). Thus, ordering the events by the proper time of the source allowed us to clearly identify two phases of the evolution of the Tohoku foreshocks.

The universal time interval between adjacent foreshocks is $T_k = t_k - t_{k-1}$, $k \geq 2$. The average value of the interval is calculated using formula $T = dt/dx$. Using formulas (4) and $<g> = T$, we obtain

$$\sigma = \frac{d}{dx} \ln T \quad (7)$$

and determine the foreshock deactivation coefficient (see Table 1).

**Table 1.** Tohoku foreshocks deactivation coefficient

| $x$ | 1 – 103 | 104 – 166 |
|---|---|---|
| $\sigma$ | 0 | 0.065 |

Let us pay attention to the constancy of the deactivation coefficient in the range from $x = 104$ to $x = 166$. The constancy of $\sigma$ indicates that the Omori law established for the main phase of aftershock evolution is fulfilled [2]. We may be dealing with a stream of aftershocks generated by a powerful impact with a magnitude of $M = 7.3$.

Let us compare Table 1 with Table 2, which shows the aftershock deactivation coefficients in the initial phase and in the main phase of source relaxation according to the data of [2].

**Table 2.** Tohoku aftershocks deactivation coefficient

| $x$ | 168 – 800 | 801 – 3000 |
|---|---|---|
| $\sigma$ | 0 | 0.0014 |

We see an interesting similarity in the change in the evolutionary regimes of foreshocks and aftershocks. In both episodes of the evolution of the source, a sudden transition from the degenerate

Omori law ($\sigma = 0$) to the classical Omori law ($\sigma = const > 0$) is observed. However, it is possible that this is characteristic only of this particular earthquake. It should be noted that the strongly reduced value of the deactivation coefficient in the main phase of aftershock evolution is consistent with the general pattern, the essence of which is that the value of $\sigma$ is lower, the higher the magnitude of the main shock [17].

### 3. Discussion

The concept of the proper time of the source has enriched the possibilities of experimental study of earthquakes. In an analysis of the Tohoku earthquake aftershocks, three phases of relaxation of the source after the main shock were previously found. In the initial phase $\sigma = 0$. In the main phase, the deactivation coefficient is a positive value that does not change over time, and the transition from the initial phase to the main phase occurs abruptly. Finally, in the recovery phase $\sigma(t)$ changes chaotically over time [2].

According to the classification proposed in [18], the Tohoku earthquake belongs to one of the six types of tectonic earthquakes. This species has been called the complete classical triad. We plan to study the above-mentioned properties of foreshocks and aftershocks on other representatives of complete classical triads, as well as on earthquakes belonging to other types of triads.

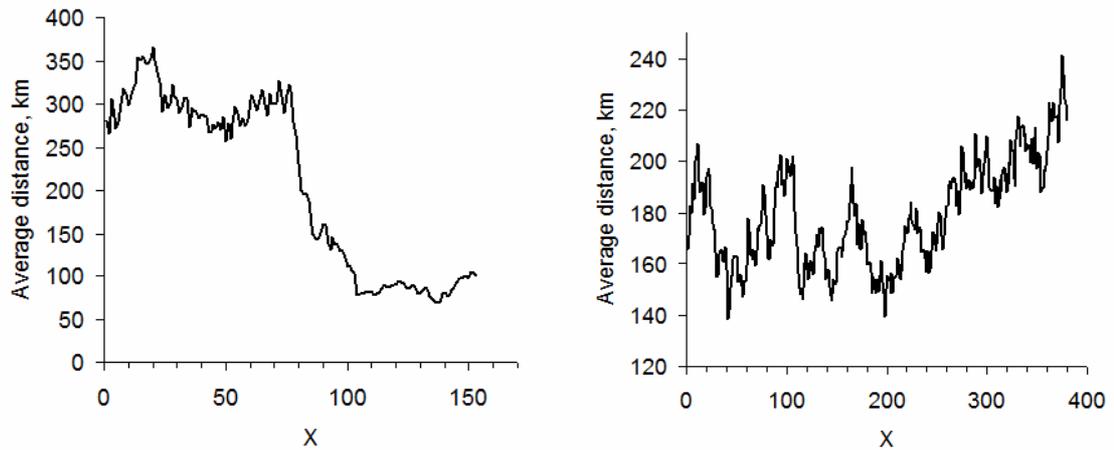

**Fig. 6**. Convergence of 166 foreshocks (left) and divergence of 400 aftershocks (right) of the Tohoku earthquake. The horizontal axis shows the source proper time. The vertical axis shows the average distance from the mainshock epicenter to the foreshock and aftershock epicenters.

In conclusion of the discussion, we will touch on the issue of the spatio-temporal distribution of earthquakes. The preliminary study consisted of introducing a circular coordinate system $(r,\varphi)$ on the earth's surface with its center at the epicenter of the main shock. The $r(x)$-coordinates of the epicenters of 166 foreshocks and 400 aftershocks are shown in Figure 6. The average values of the distance from the epicenter of the main shock obtained by moving average over 20 points. We see a trend towards decreasing distance for foreshocks and increasing distance for aftershocks. The phenomenon was discovered earlier through statistical studies of earthquakes [5, 19] and was called foreshock convergence and aftershock divergence. In fairness, it should be said that the convergence of foreshocks is probably a manifestation of the effect of the contraction of hypocenters of relatively weak earthquakes to the plane of the future main rupture [20].

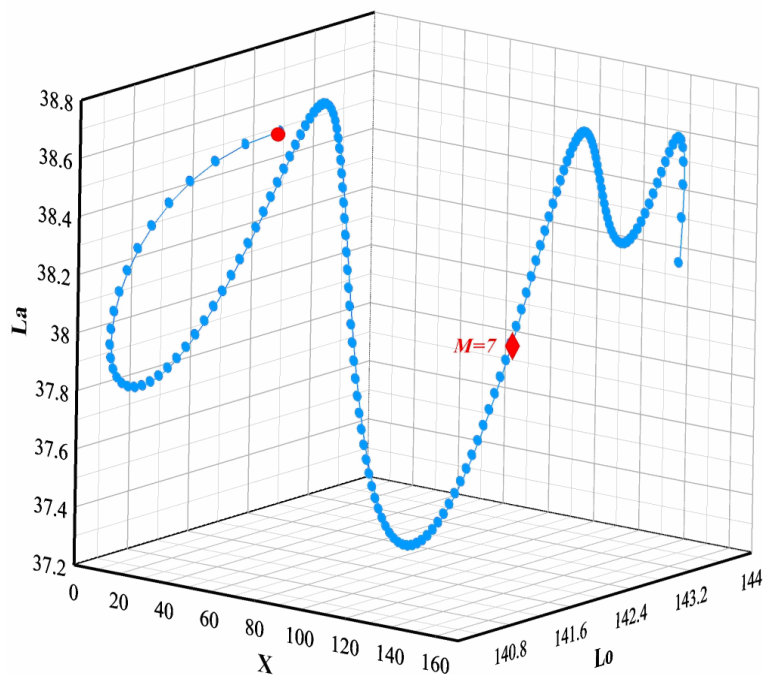

**Fig. 7**. Movement of foreshock epicenters ordered by proper time of the Tohoku earthquake source.

And finally, we present in Figure 7 the foreshock epicenters depending on proper time. The overall picture resembles a string of pearls. Allowing for some laxity, we can imagine the flow of foreshocks as a tracer visualizing the flow of Umov energy in a stress-strained rock mass. Let's pay attention to the curvature and torsion of the curve along which the epicenters are aligned. Our own interpretation, right or wrong, is that we have discovered a vortex movement of the Umov energy flow.

## 4. Conclusion

We investigated the foreshocks of the Tohoku earthquake using the concept of proper time, similar to what we did in our study of aftershocks [2]. A peculiar relativism of time, which, however, has no direct relation to the theory of relativity, manifested itself in the fact that the flow of the proper time of the source does not coincide with the flow of world time. Ordering the foreshock and aftershock sequences by proper time allowed us to identify interesting properties of the source that were not previously discovered when ordering events by universal time. We assume that the unevenness of the flow of proper time relative to universal time indicates the non-stationarity of the source parameters.

*Acknowledgments*. We express our sincere gratitude to B.I. Klain and A.D. Zavyalov for fruitful discussions and support. We thank colleagues at the US Geological Survey for lending us their earthquake catalogs USGS/NEIC for use. The work was carried out within the framework of the planned tasks of the Ministry of Science and Higher Education of the Russian Federation to the Institute of Physics of the Earth of the Russian Academy of Sciences.